\documentclass[11pt]{article}
\pdfoutput=1

\usepackage{jheppubmod}
\usepackage{hyperref}

 \usepackage{mathtools}
\usepackage{psfrag}
\usepackage{array}
\usepackage{amssymb}
\usepackage{amsmath}
\usepackage{cancel}

\usepackage{titlesec}

\titleformat{\paragraph}
{\normalfont\normalsize\bfseries}{\theparagraph}{1em}{}
\titlespacing*{\paragraph}
{0pt}{3.25ex plus 1ex minus .2ex}{1.5ex plus .2ex}

\usepackage{amsthm}
\usepackage{graphicx}
\usepackage{caption}
\usepackage{paralist}
\usepackage{comment}

\usepackage{epsfig}

\def\pb[#1,#2]{\{#1, #2\}}
\def\deb[#1,#2]{[#1,#2]_{\text{D.B.}}}

\def\Or[#1]{{\text{O}}\left({#1}\right)}
\def\dotl[#1,#2]{\left\langle #1,\, #2 \right\rangle}
\def\dotlb[#1,#2]{\left\langle #1,\, #2 \right\rangle}
\def\dotlm[#1,#2]{\left[ #1,\, #2 \right]}
\def\dotp[#1,#2]{(\vect{#1} \cdot\vect{#2})}
\def\aff[#1,#2]{\hat{#1}(#2)}

\def\n4sym{{\cal N}=4 SYM}
\def\>{\rangle}
\def\<{\langle}
\def\weight[#1,#2,#3]{\{(#1),#2,#3\}}
\def\ads[#1]{$\text{AdS}_{#1}$}

\newcommand{\be}{\begin{equation}}
\newcommand{\ee}{\end{equation}}
\newcommand{\ba}{\begin{align}}
\newcommand{\ea}{\end{align}}
\newcommand{\bs}{\begin{split}}
\def\sess\end{split}
\newcommand{\vect}[1]{{\boldsymbol{#1}}}

\title{Elliptic genera from multi-centers}
\author[]{Nava Gaddam,}
\affiliation[]{Institute for Theoretical Physics and Center for Extreme Matter and Emergent Phenomena,
Utrecht University, 3508 TD Utrecht, The Netherlands.}
\emailAdd{gaddam@uu.nl}

\date{}

\abstract{I show how elliptic genera for various Calabi-Yau threefolds may be understood from supergravity localization using the quantization of the phase space of certain multi-center configurations. I present a simple procedure that allows for the enumeration of all multi-center configurations contributing to the polar sector of the elliptic genera\textemdash explicitly verifying this in the cases of the quintic in $\mathbb{P}^4$, the sextic in $\mathbb{WP}_{(2,1,1,1,1)}$, the octic in $\mathbb{WP}_{(4,1,1,1,1)}$ and the dectic in $\mathbb{WP}_{(5,2,1,1,1)}$. With an input of the corresponding `single-center' indices (Donaldson-Thomas invariants), the polar terms have been known to determine the elliptic genera completely. I argue that this multi-center approach to the low-lying spectrum of the elliptic genera is a stepping stone towards an understanding of the exact microscopic states that contribute to supersymmetric single center black hole entropy in $\mathcal{N}=2$ supergravity.}
\begin{document}
\maketitle
\section{Introduction}
Since the seminal work of \cite{Strominger:1996sh,Maldacena:1997de}, the study of black hole microscopics has received significant attention. A quantum understanding of black holes had been plagued with several problems for decades. Of them, the apparent infiniteness of the Hilbert space of states associated to the horizons was particularly striking \cite{'tHooft:1984re}. The microscopic explanation of black hole entropy elegantly solved this problem in a naturally UV complete setting; that of string theory. \\

At this juncture, one had two obvious paths to deliberate between. One, to take the finiteness of the space of states as a final result from string theory and seek an understanding of the special nature of interaction between these degrees of freedom that endow black holes with their exceedingly mysterious dynamics; this is perhaps an obvious path leading towards a quantum  understanding of gravitational dynamics. The other, perhaps more modest path, would have been to first seek a refined understanding of the static; more than a mere count of states, that is. Both paths have been travelled, even extensively if one might add, and yet it is fair to say that much is left to be understood. The microscopic counting of \cite{Strominger:1996sh,Maldacena:1997de} accounts for the number of states, to leading order in charges, that yield black hole entropy. What is counted on the microscopic front is an index\textemdash a sum over all `angular momentum states'. The macroscopic black hole is a singlet in that it is a static, stationary spherically symmetric solution to the bulk supergravity equations of motion. While the leading order counting of states matches with the macroscopic entropy, an often under-appreciated problem is the lack of understanding of what each of these states is. One reason for the difficulty in identifying them exactly is that a sum over states of a given representation under the angular momentum group is not a protected quantity. On the macroscopic front, however, a sum over various black hole configurations may seem unnatural. In examples with sufficient amount of supersymmetry, significant progress has been made \cite{Sen:2007qy,Dabholkar:2010rm,Dabholkar:2004yr,Dabholkar:2005dt,Dabholkar:2010uh,Dabholkar:2014ema,Dabholkar:2012nd,Murthy:2015zzy}. Nevertheless, in cases with lesser supersymmetry, the picture is a lot less clear. In this article, I study one such set-up that is least understood in this context\textemdash the one of \cite{Maldacena:1997de}. The macroscopic black holes of interest here are supersymmetric dyonic ones in $\mathcal{N}=2$, $d=4$ supergravity obtained from a Calabi-Yau\footnote{In this article, by a Calabi-Yau manifold, I refer to one with maximal holonomy. More is known about this setup when one relaxes this condition \cite{Dabholkar:2010rm}.} compactification of M-Theory (an obvious, equivalent picture exists in the Type IIA setting). The microscopic states that count their entropy arise in what has been dubbed the MSW CFT, after the authors of \cite{Maldacena:1997de}. \\

The MSW CFT is a (0,4) supersymmetric non-linear sigma model that is believed to flow to a conformal fixed point in the IR. The supersymmetric states of this theory can be counted via an index\textemdash the modified elliptic genus \cite{deBoer:2006vg,Gaiotto:2006wm}
\begin{equation}\label{eqn:modellgen}
\mathcal{Z}\left(q,\bar{q},\tilde{y}\right) ~ = ~ \mathrm{Tr}\left(\dfrac{1}{2} F^2 \left(-1\right)^F q^{L_0 - \frac{c_L}{24}} \bar{q}^{\bar{L}_0 - \frac{c_R}{24}} \tilde{y}^{2 J}\right) \, ,
\end{equation}
where $q = e^{2\pi i \tau}$ with $\tau$ being the modulus of the torus on which the theory is to live and $\bar{q}$ is its complex conjugate. $F$ refers to the fermion number as in the case of the standard Witten index. $c_L$ and $c_R$ label the left and right moving central charges of the field theory. Finally, $\tilde{y}=e^{2\pi i z}$ is a fugacity associated to the elliptic variable $z$ and is raised with a `chemical potential' $J$ associated to the eigenvalue of the generator\footnote{Note that this $U(1)$ generator is to be distinguished from $J_3$, appearing in the next sections. The latter refers to the angular momentum of the macroscopic black hole dual to a given state in the field theory. I thank the anonymous JHEP referee for pointing out this important disambiguation from a previous version of this preprint.} of the right-moving $U(1)$ algebra arising from the self-dual part of the $(1,1)$ forms of the threefold. BPS excitations of this theory, counted by the above partition function, have been shown to grow\textemdash to leading order in charges\textemdash exactly as does the entropy of a macroscopic single center supersymmetric black hole in the four dimensional $\mathcal{N}=2$ supergravity theory \cite{Maldacena:1997de}. Exciting as that result may already be, the above modified elliptic genus \eqref{eqn:modellgen} in fact enjoys an even richer structure. It is a \textit{weak Jacobi form} of weight $(-\frac{3}{2},\frac{1}{2})$ and is endowed with a $\Theta$-decomposition\footnote{$\Theta_\gamma$ arises in a decomposition of the modular invariant theta function associated to the flux lattice of the Calabi-Yau being compactified on. For details, see \cite{deBoer:2006vg,Gaiotto:2006wm}.} in terms of vector valued modular forms $\mathcal{Z}_\gamma$ as \cite{deBoer:2006vg,Gaiotto:2006wm}
\begin{equation}
\mathcal{Z}\left(q,\bar{q},\tilde{y}\right) ~ = ~ \sum_{\gamma = 0}^n \mathcal{Z}_\gamma\left(q\right) \Theta_\gamma \left(q,\bar{q},\tilde{y}\right) \, ,
\end{equation}
where $\gamma$ labels the independent elements of the corresponding discriminant group. Loosely speaking, the vector valued modular form $\mathcal{Z}_\gamma$ captures the growth of states of the partition function \eqref{eqn:modellgen} while the $\Theta_\gamma$ functions\textemdash forming modular representations of weight $(\frac{1}{2}h^{(1,1)}(CY_3) -1, \frac{1}{2})$\textemdash add to the rich pole structure of the modified elliptic genus. While much more can be said of this decomposition than is within the scope of this article or my current understanding, I will restrict my attention to the vector $\mathcal{Z}_\gamma$ which captures the growth of states that endow the macroscopic black holes with their entropy. For simplicity, I will also consider those compactifications with $h^{(1,1)} = 1$; this allows for a study of uni-modulus supergravity theory on the macroscopic front. Furthermore, given that the $\Theta_\gamma$ functions are then of weight $(0,\frac{1}{2})$, $\mathcal{Z}_\gamma$ would carry modular weight $-\frac{3}{2}$. Finally, $\mathcal{Z}_\gamma$ is also endowed with a $q$-expansion\textemdash the coefficients of which capture a sum over all the black hole microstate degeneracies falling in various representations of the space-time angular momentum\textemdash that begins with a negative power of $q$. The polar sector of the modular form is defined to be the set of all terms in this expansion with negative powers of $q$; knowledge of all the polar terms uniquely determines the entire modified elliptic genus \cite{deBoer:2006vg,Gaiotto:2006wm,Gaiotto:2007cd}. \\

The leading order growth of coefficients in this $q$-expansion of $\mathcal{Z}_\gamma$ is what a Cardy-estimate of the growth of states counts. However, as the trace in the definition of the index indicates, all bound states with a total charge equalling that of a single center black hole also contribute to the corresponding term in the $q$-expansion. While contributions from any one of these bound states may be small, the number of possible configurations clearly grows as the number of partitions of the charge/energy level in question. As Ramanujan famously showed, this number grows exponentially; much like the Cardy estimate, one might observe. This raises the following question\textendash
\begin{center}
\textit{What states is the Cardy formula really counting?}
\end{center}

It is the aim of this article to provide a disambiguation of this issue and work towards an answer to the above question. Ideally, a clinching answer would be a listing of all bound states contributing to a large charge coefficient in the $q$-expansion of $\mathcal{Z}_\gamma$ leaving an appropriate single-center entropy and the origin of the corresponding states behind. However, the exponentially large number of such bound states renders this practically impossible to achieve. One hopes to uncover a structure in the contributions arising from these bound states that may be extrapolated to arbitrary charges. Since the polar terms are the low-lying states and are those that actually entirely determine the modular form uniquely, one may imagine that they provide for a good starting point. \\ 

One may in fact opt for a more direct approach to understand single-center black hole entropy: it has been shown \cite{Bringmann:2013yta} that sub-leading corrections to the growth of states of the modified elliptic genus depend on their representation of angular momentum. It is certainly an interesting way forward and deserves more attention than it has received, in my opinion. Notwithstanding this aside, I take the former approach in this article. \\

In this article, I present a systematic way to identify all the multi-center configurations that enumerate the polar states of the vector valued modular form $\mathcal{Z}_\gamma$ using the equivariant refined index introduced in \cite{Manschot:2010qz,Manschot:2011xc}; as has been noted before \cite{Denef:2007vg} no single-center configurations contribute to the polar sector. Along the way I find some interesting results regarding the existence\textemdash or lack thereof\textemdash of certain three-center configurations involving $D2(\bar{D2})$ charges. I work by example to identify all the multi-centers needed to uniquely determine the elliptic genera of the following Calabi-Yau threefolds: the quintic in $\mathbb{P}^4$, the sextic in $\mathbb{WP}_{(2,1,1,1,1)}$, the octic in $\mathbb{WP}_{(4,1,1,1,1)}$ and the dectic in $\mathbb{WP}_{(5,2,1,1,1)}$. All results in this article that have been derived before in \cite{Gaiotto:2006wm,Gaiotto:2007cd}, agree with those references; furthermore, as in the said references, I use the known Gopakumar-Vafa invariants. These were computed in \cite{Katz:1999xq,Klemm:1992tx,Huang:2006hq} while the relation of Gromov-Witten invariants to Gopakumar-Vafa invariants is excellently reviewed in \cite{Hori:2003ic}\footnote{See chapters 33 and 34, in particular.}. Finally, the equivalence of these to Donaldson-Thomas invariants was conjectured and proved in \cite{COM:501896,COM:501908,JL,BL}. \\

The rest of this article is organized as follows. In Section \ref{sec:theindex}, I review the relevant multi-center configurations of interest and provide an intuitive argument for what the appropriate index that counts their interaction degrees of freedom must be; furthermore, I also spell out the prescription to be used to identify those configurations that contribute to the polar terms of the elliptic genera under consideration. In Section \ref{sec:m5ellgenera}, I explicitly compute the said indices for several examples. I conclude with a discussion in Section \ref{sec:discussion} 

\section{The refined equivariant index}\label{sec:theindex}
In this section, I will first review the phase space of multi-center configurations, merely stating results and known facts. Details may be found in \cite{Denef:2007vg,Denef:2000nb,Bates:2003vx}. I then move on to a present an intuitive explanation for the appropriate index that counts multi-center degeneracies. \\

Multi-center configurations in $\mathcal{N}=2$ supergravity are characterized by a metric ansatz for stationary solutions
\begin{equation}
ds^2 ~ = ~ - e^{2 U(\vec{r})} \left(dt + a(\vec{r})\right)^2 + e^{-2U(\vec{r})} d\vec{r}^2 \, ,
\end{equation}
with $a(\vec{r})$ denoting a Kaluza-Klein one-form and $U(\vec{r})$ the scale factor. The scalars in the vector multiplet are typically called $t^a$, with the index $a$ running over the set of all vector multiplets. Since I restrict to Type IIA compactifications with $h^{(1,1)} = 1$, there is only one modulus in the theory allowing for a dropping of the index $a$. The real and imaginary decomposition of the modulus is labelled as $t = B + i \mathrm{J}$. Denoting the charge lattice by $\Gamma$, a given center carries charges that form a vector $\alpha \in \Gamma$; for the case at hand in uni-modulus supergravity, this vector is four-dimensional: $\left(p^0, p, q, q_0\right)$. The charges $p^0$ and $p$ are magnetic in my conventions and correspond to $D6$ and $D4$ brane charges in Type IIA language. Whilst $q$ and $q_0$ are electric charges corresponding to $D2$ and $D0$ excitations. There is a natural symplectic inner product between two such charge vectors $\alpha$ and $\tilde{\alpha}$
\begin{equation}\label{eqn:symplprod}
\langle \alpha , \alpha^\prime\rangle ~ = ~ q_0 p^{\prime 0} + q p^\prime - q^\prime p - q^\prime_0 p^0
\end{equation}
and it is clearly antisymmetric. For a multi-center configuration with total charge $\gamma = \sum_i \alpha_i$, with each center at a location $\vec{r}_i$, the scale factor and the value of the modulus $t$ are uniquely fixed by the `attractor equations'\footnote{I spell out the exact quantities appearing in these equations in the ensuing page.} \cite{Denef:2000nb}
\begin{align}\label{eqn:attractoreqns}
- 2 e^{-U(\vec{r})} \mathrm{Im} \left[e^{-i \phi} \Omega(t(\vec{r}))\right] ~ &= ~ \beta + \sum_{i=1}^{n} \dfrac{\alpha_i}{\left|\vec{r} - \vec{r}_i\right|} \qquad \text{with} \nonumber \\
\phi ~ &= ~ \arg\left(Z_\gamma\right) \, .
\end{align}
The moduli space of the scalars in the vector multiplet is a special K\"{a}hler manifold that has a principal bundle over its base space with a structure group $Sp(2n_v +2)$, where $n_v$ is the number of vector multiplets in the theory. Calling the coordinates on the fibers of the appropriate vector bundle $X^A$ and $F_A$, the manifold affords a nowhere vanishing holomorphic symplectic section. The index $A$ runs over $n_v +1$ indices; therefore $A \in \{0,1\}$. Now, in the above attractor equations, $\Omega(t(\vec{r})) = - e^{\mathcal{K}/2} \left(X^A, F_A\right)$ is the said symplectic section. $\mathcal{K} = - \ln \left[i \left(F_A \bar{X}^A - \bar{F}_A X^A\right)\right]$ is the K\"{a}hler potential associated to $\mathcal{M}_v$. Furthermore, $\beta$ is a constant vector given in terms of the asymptotic value $t_\infty$ of the modulus by 
\begin{equation}
\beta ~ = ~ - 2 \mathrm{Im} \left[e^{-i\phi} \Omega\left(t_\infty\right)\right] \, .
\end{equation}
In the one-modulus supergravity theory at hand, projective symmetry allows for a fixing of the $X^0$ coordinate to unity leaving the only modulus $t = X^1/X^0$. The coordinates $F_A$ on the fibers are in fact derived, as $F_A = \partial_A F$, from the prepotential $F$ of the theory:
\begin{equation}\label{eqn:prepotential}
F\left(X^0, X^1\right) ~ = ~ - \dfrac{k}{6} \dfrac{\left(X^1\right)^3}{X^0} + \dfrac{\mathcal{A}}{2} \left(X^1\right)^2 + \dfrac{c_2 \cdot P}{24} X^0 X^1 + \text{instantons} \, .
\end{equation}
Here, I work in the following normalizations
\begin{equation}\label{eqn:CYnormalizations}
\int_{CY_3} ~ = ~ \omega \wedge \omega \wedge \omega \, ; \, \int_{CY_3} \omega \wedge c_2 (CY_3) ~ = ~ c_2 \cdot P \, ; \, \int_{CY_3} \omega_a \wedge \omega^b ~ = ~ \delta^b_a \, ; \, \int_{CY_3} \omega^c ~ = ~ 1 \, ,
\end{equation}
where the $\omega$ form a basis of integer two-cycles in the threefold. Finally, the half-integer constant in the quadratic piece of the prepotential is given by $\mathcal{A} = k/2$ mod $1$. For the purposes of this article, the instanton corrections may be ignored\footnote{While one may be worried about the validity of the supergravity regime\textemdash without instanton corrections\textemdash in the case of small charge configurations, they turn out to have a rather specific and easily controlled effect insofar as regularity of solutions is concerned. I will be explicit about this effect in further sections.}. Finally, the one-form $a\left(\vec{r}\right)$ is determined in terms of the Hodge-star operator of the three flat dimensions by
\begin{equation}
\star_3 \mathrm{d} a\left(\vec{r}\right) ~ = ~ \left\langle d \sum_{i = 1}^{n} \dfrac{\alpha_i}{\left|\vec{r} - \vec{r}_i\right|}, \beta + \sum_{i = 1}^{n} \dfrac{\alpha_i}{\left|\vec{r} - \vec{r}_i\right|} \right\rangle \, .
\end{equation}
In what follows, I shall label $\left|\vec{r} - \vec{r}_i\right|$ by $r_{ij}$ and $\langle\alpha_i,\alpha_j\rangle$ by $\alpha_{ij}$. The `integrability equations' 
\begin{equation}\label{eqn:Denefeqns}
\sum_{\substack{j=1 \\ j \neq i}}^{n} \dfrac{\alpha_{ij}}{r_{ij}} ~ = ~ c_i \qquad \text{with} \qquad c_i ~ = ~ 2 \mathrm{Im} \left[e^{-i \phi} Z_{\alpha_i}\right]
\end{equation}
ensure the existence of an $a\left(\vec{r}\right)$ such that the configuration is supersymmetric. Finally, the central charge $Z_\gamma$ is given by 
\begin{align}\label{eqn:centralcharge}
Z_\gamma ~ &= ~ \langle \gamma, \Omega\left(t\right)\rangle \nonumber \\
&= ~ e^{\mathcal{K}/2} \left[p^A F_A - q_A X^A\right] \nonumber \\
&= ~ e^{\mathcal{K}/2} X^0 \left[\dfrac{k}{6} p^0 t^3 - \dfrac{k}{2} p t^2 - \tilde{q} t - \tilde{q}_0\right] \, ,
\end{align}
where the charges have been written with a tilde suggestively, to indicate that they are not integer quantized. The exact quantization can be spelled out and I shall do so in Section \ref{sec:m5ellgenera}. Furthermore, the K\"{a}hler potential can be computed from its definition as 
\begin{align}\label{eqn:expkahlerpot}
 e^{-\mathcal{K}} ~ &= ~ i\left( \bar{X}^AF_A - X^A\bar{F}_A \right) \nonumber \\
 &= \dfrac{4}{3}k\mathrm{J}^3 |X^0|^2 \, .
\end{align}
Having specified all the quantities appearing in the attractor equations \eqref{eqn:attractoreqns}, there is one additional and extremely important constraint that these multi-center configurations must satisfy; that of regularity. One might impose this by demanding the positivity of the scale factor in front of the $d\vec{r}^2$ term in the metric. The attractor equations can be shown to imply that\textemdash for a configuration with $i$ centers located at $\vec{r}_i$\textemdash this is equivalent to evaluating the entropy on the regularity vector appearing on the right hand side of the attractor equations \eqref{eqn:attractoreqns} \cite{Bates:2003vx}
\begin{equation}\label{eqn:multiregularity}
S\left(\beta + \sum_{i=1}^{n} \dfrac{\alpha_i}{\left|\vec{r} - \vec{r}_i\right|}\right) ~ > ~ 0 \, , \quad \forall \quad \vec{r} \in \mathbb{R}^3 \, .
\end{equation}
One may in fact solve for the attractor equations in full generality in uni-modulus supergravity to spell out this entropy function explicitly \cite{Shmakova:1996nz}. I present the explicit formula in the next section. \\

At this stage, however, the goal is to understand how one may calculate the total number of degrees of freedom associated to such a gravitational solution. To this end, an `equivariant refined index' for such bound states as described was proposed in \cite{Manschot:2010qz,Manschot:2011xc}. While I leave the technical derivation of this refined index to those papers, in what follows I argue for the correctness of their proposed index. It is my hope that this discussion gives an intuitive picture leaving the more rigorous, technical treatment to those original papers. \\

Consider the solutions of the integrability equations \eqref{eqn:Denefeqns}. Although the equations are seemingly simple, they are deceptively so. There is no general analytic solution set to these equations. However, for a given configuration, one might numerically solve for the positions $r_i$ of the black hole centers. In general, there is a non-trivial angular momentum associated to every point in the space generated by the solutions of the Denef equations; for a single-center on the other hand, spherical symmetry ensures that this angular momentum is zero. There is also an action of the rotation group $SO(3)$ that leaves the space of solutions invariant; this is just a rotation of the whole configuration of the bound state in space-time. The corresponding study of such spaces, with an action of a group, in the Mathematics literature is that of Hamiltonian spaces and equivariant cohomology. Leaving the intricate details to the excellent review \cite{Vergne}, I will resort to a more sketchy and qualitative consideration to tell the number of degrees of freedom to be associated to such bound states. While a two center solution can immediately be imagined, increasing the number of centers in the problem prevents easy visualization. For instance, the integrability equations for a two center problem essentially fix the distance between the two centers\footnote{Up to translations that can be gauged by fixing one of the centers to be at the origin.}. Rotating this configuration in space-time generates a round sphere as the space of solutions; the sphere is clearly smooth and symplectic. To generalize this to phase spaces of solutions of a configuration with higher number of centers is an open problem in Mathematics. Nevertheless, one can write down a symplectic two-form on the phase space of solutions of the integrability equations \cite{deBoer:2008zn}. It is again a non-trivial task to prove that a given two-form is indeed non-degenerate on the phase space. Therefore, that the phase space is symplectic is best left to be conjectural at this juncture. This phase space is classical. An `equivariant volume element' of this phase space (read as a volume element that accounts for the non-trivial angular momentum at each point in the space) is one that accounts for the interaction between the black hole centers. The phase space is built out of these equivariant volume elements. This is an extremely important insight. It tells us, among other things, that quantizing this phase space yields a quantum index of the interaction between the black hole centers \cite{Manschot:2011xc}. Such a quantum index is to keep track of the interaction degrees of freedom of the black holes. While this is a very naive picture, a more rigorous discussion can be found in \cite{Manschot:2011xc}. In the following, I will take a slightly different perspective from \cite{Manschot:2011xc} to understand this index. \\

In mathematical terms, quantization of a phase space that is symplectic, is best understood with the theory of \emph{Geometric Quantization}. An excellent review for aspects relevant to us can be found in \cite{Ritter}. The basic idea is the following - given a line bundle (called the pre-quantum line bundle) and a space of sections of this line bundle (called the pre-quantum space) on the phase space, one can construct a quantum space as a set of subspace of sections of this pre-quantum line bundle that vanishes under the action of a covariant derivative that is defined on the line bundle (via the corresponding connection). Physically speaking, a pre-quantum space can be identified with the space of square integrable sections on an appropriate pre-quantum line bundle. These sections would, upon quantization, build up the quantum space - the Hilbert space of states. In the setting at hand, apart from square integrable sections on the line bundle, we also have a spinor bundle consisting of sections corresponding to the fermionic supersymmetry generators in the theory. A clever ploy would be to choose the covariant derivative to be the Dirac operator on the phase space. This is a clever choice for the formally defined equivariant index of the Dirac operator now counts the quantum states in the theory. This is a direct consequence of the definition of the index of the Dirac operator. It is worth understanding this index better for this is what is to be computed, eventually. \\

Given a vector bundle $E \to M$ on a manifold $M$ with an action of a group $G$ acting on it; consider the action of the group on $M$ such that it lifts to an action on $E$. The Dirac operator (whose action is assumed to commute with $G$ henceforth) is now defined on the space of sections of this vector bundle as
\begin{equation}
 D \colon \Gamma(E) \longrightarrow \Gamma(E).
\end{equation}
By definition, the equivariant index of this Dirac operator, for an element $g\in G$, is
\begin{equation}
 \mathrm{Ind}_G(g, D) = \mathrm{Tr}_{\mathrm{Ker} D^+}(g) - \mathrm{Tr}_{\mathrm{Ker} D^-}(g).
\end{equation}
Equivalently, considering the Lie Algebra $\mathfrak{g}$ of $G$ and an element $x=\ln(g) \in \mathfrak{g}$, the equivariant index can be defined as \cite{BerlineVergne}
\begin{equation}
 \mathrm{Ind}_G(\mathrm{exp}(x), D) = \dfrac{1}{(2\pi i)^\frac{n}{2}} \displaystyle\int_M Ch_\mathfrak{g}(x, E)\hat{A}_\mathfrak{g}(x, M),
\end{equation}
where $Ch$ denotes the Chern character and $\hat{A}$ denotes the usual $A$-roof genus; this is also called Kirilov's formula. For the purposes of this article, in the spirit of the Witten index, picking an element $y^{2J_3} \in G$, where $y$ is a formal generating parameter and $J_3$ is the third generator of the angular momentum algebra of the rotations in space-time, the index can now be written as\footnote{A dependence on the complexified K\"{a}hler parameter $t$ is implicit if one is to work globally in the moduli space; locally, however, the index is constant.}
\begin{equation}
 g_{\text{ref}}(\{ \alpha_i \}, y) = \mathrm{Tr}_{\mathrm{Ker} D^+}\left[(-y)^{2J_3}\right] - \mathrm{Tr}_{\mathrm{Ker} D^-}\left[(-y)^{2J_3}\right].
\end{equation}
This is the index for a configuration of black hole centers carrying charges $\alpha_i$ that form a bound state satisfying the integrability equations \eqref{eqn:Denefeqns}. $g_{\text{ref}}$ stands for the \emph{refined} index; to avoid confusion, I merely stick to conventional notation used in \cite{Manschot:2011xc}. This can further be shown to reduce to \cite{Manschot:2011xc}
\begin{equation}
 g_{\text{ref}}(\{ \alpha_i \}, y) = \int_{\mathcal{M}_n} Ch(\nu, \mathcal{L}) \hat{A}(\nu, \mathcal{M}_n),
\end{equation}
where $\mathcal{L}$ is the line bundle, $\mathcal{M}_n$ is the phase space of an $n$-centered problem solving the integrability equations and $\nu = \ln y$.\\

The idea now, is to compute this index via Localization. Knowing the group action on the phase space, a localization technique under an Abelian subgroup of this group ($U(1)$ of $SO(3)$) results in a localization of the black hole centers along a line, say the $z$ axis, with manifest $U(1)$ symmetry; the symmetry being rotations about the axis of localization. This renders a non-vanishing contribution to the index only from the fixed points that are the black hole centers. What was originally a problem in $\mathbb{R}^3$ has now localized to a problem on a line with the centers lying at positions, say ${z_i}$. With this knowledge, one may write down a `superpotential' whose fixed points are given by exactly the fixed points of localization \cite{Manschot:2011xc}
\begin{equation}\label{eqn:superpot}
\hat{W}\left(\lambda, \{z_i\}\right) ~ = ~ - \sum_{i<j} \alpha_{ij} \mathrm{sign} \left[z_j - z_i\right] \ln \left|z_j - z_i\right| - \sum_i \left(c_i - \dfrac{\lambda}{n}\right) z_i \, .
\end{equation}
This superpotential is a function of $n+1$ variables: the $n$ centers and a parameter $\lambda$. With these considerations, the index can now be written in its computationally easiest form as
\begin{equation}\label{eqn:gref}
 g_{\text{ref}}(\{ \alpha_i \}, y) = \dfrac{(-1)^{\sum_{i < j} \alpha_{ij}+n-1}}{\left(y-y^{-1}\right)^{n-1}} \displaystyle\sum_{p} s(p) ~ y^{\sum_{i < j}\alpha_{ij}\mathrm{sign}[z_j -z_i]},
\end{equation}
where $p$ corresponds to a given regular configuration of black hole centers that satisfy the integravility equations and $s(p) = - \mathrm{sign} \det \hat{M}$, with $\hat{M}$ being the Hessian of $\hat{W}(\lambda, \{z_i\})$ with respect to $z_1, \dots , z_n$. Upon specifying $y=-1$, this $g_{ref}$ is exactly that quantum index which computes the interaction degrees of freedom arising from a given multi center black hole solution to supergravity. Another interpretation of this quantity is that of the Poincar\'{e} polynomial associated to the moduli space of the quiver representations: each center in the configuration arises from a D-brane that may be treated as a node with an Abelian gauge group associated to it. With bifundamentals extending between the bound centers playing the arrows, these configurations do indeed take the guise of a quiver diagram\cite{Denef:2002ru}. Topological invariants associated to the moduli space of representations of these quivers have been shown to be enumerated by this index $g_{\text{ref}}$ \cite{Manschot:2013sya,Manschot:2014fua}. With the knowledge of the interaction degrees of freedom between the black hole centers, the total degeneracy associated to a multi-center black hole configuration can now be naturally written as
\begin{equation}\label{eqn:finalindex}
\bar \Omega(\{\alpha_i\};t) = \frac{g_{ref}(\{\alpha_i\};t)}{|{\rm Aut}(\{\alpha_i\})|} \prod\nolimits_{i=1}^n \bar \Omega^S(\alpha_i),
\end{equation}
where $\bar \Omega(\{\alpha_i\};t)$ is the total degeneracy associated to the multi-center configuration in question, $\bar \Omega^S(\alpha_i)$ corresponds to the rational index associated to a single black hole center carrying charge $\alpha_i$ and the $|\mathrm{Aut}(\{\alpha_i\})|$ factor\footnote{$|{\rm Aut}(\{\alpha_i\})| = \prod_k z_k!$.} takes repeated centers into account. The single center indices are input parameters. These rational indices are given, in terms of the integer invariants, by
\begin{equation}
 \bar \Omega^S(\alpha_i) = \displaystyle\sum_{m|\alpha_i} m^{-1} \dfrac{y - y^{-1}}{y^m-y^{-m}} \Omega^S(\alpha_i),
\end{equation}
where $\Omega^S(\alpha_i)$ are the integer invariants of the single centers. It may be noted that the product of these rational indices is the mathematical counterpart of the symmetric product of the moduli spaces in the string regime, that contains several singularities \cite{Gaiotto:2006wm,Gaiotto:2007cd}. From a supergravity perspective, however, this product can physically be understood as arising from the Bose-Fermi statistics of the interacting single center black holes \cite{Manschot:2010qz}. This essentially negates all troubles encountered with singularities in the geometric counting.\\

Finally, a word on the regime of validity of this approach is in order. Owing to the attractor mechanism in four dimensional $\mathcal{N}=2$ supergravity theories, as the size of the modulus approaches the attractor value, it is fixed by the charges of the single center black hole towards which the modulus is being attracted. In a multi-center configuration however, bound states exist only at large values of the modulus. This is because at smaller values, one is attracted to the basin of attractor of one of the bound state constituents, owing to the attractor mechanism. Therefore, the analysis of multi-center configurations in this paper is done in the large volume limit: $\mathrm{J} \gg B$.

\section{M5-brane elliptic genera from multi-centers}\label{sec:m5ellgenera}
Having\textemdash at least morally\textemdash justified the index that computes the interaction degrees of freedom, in this section I will show how one may identify those mutli-centers that contribute to the polar terms of the MSW elliptic genus. Working by example, I explicitly show that all polar terms of the quintic in $\mathbb{P}^4$, the sextic in $\mathbb{WP}_{(2,1,1,1,1)}$, the octic in $\mathbb{WP}_{(4,1,1,1,1)}$ and the dectic in $\mathbb{WP}_{(5,2,1,1,1)}$ can be reconstructed with this approach. In the next section, will finish with an argument why this approach is well suited to identifying single-center black hole entropy in the non-polar sector. \\

To this end, one first needs to identify what the charges of individual terms of the $q$-expansion of $\mathcal{Z}_\gamma$ must be. Knowing that these charges arise from a $D4-D2-D0$ brane construction, the central charge $Z_\gamma(t)$ provides an easy tool for this purpose. $D_p$ branes often support lower dimensional brane charges. A pure $D4$ brane, for instance, supports non-zero $D2$ and $D0$ fluxes \cite{deBoer:2006vg,Gaiotto:2006wm} to cancel the Freed-Witten anomaly \cite{Freed:1999vc}. Since these branes must form localized objects as black holes in the four dimensional non-compact space in the low energy theory, their extension is entirely confined to the compact Calabi-Yau space. Every Calabi-Yau threefold has a non-vanishing structure-sheaf. Since the $D6$ brane must extend entirely in the threefold, one may view it as the structure-sheaf of the manifold and consequently, there is always one at our disposal. The central charge of a BPS brane is given by the same formula \eqref{eqn:centralcharge} as in the supergravity theory. However, the lower dimensional fluxes on the D6 brane induce additional curvature. In addition, if the brane has a non-trivial gauge bundle turned on, the charge vector of the brane would arise from turning on the relevant Chern classes. Taking all of these into consideration, the central charge takes the form \cite{Denef:2007vg}
\begin{equation}
Z_\gamma (t) ~ = ~ - \int_{CY_3} e^{U_1 + U_2 + U_3} \wedge e^{- t \omega} \wedge \left(1 + \dfrac{c_2 \left(CY_3\right)}{24}\right) \, 
\end{equation}
where the $U_i$ represent integer classes in which the Chern classes of the gauge bundle have been expanded as $c_1 = U_1 \omega$, $c_2 = U_2 \tilde{\omega}^b$ and $c_3 = U_3 \tilde{\omega}^c$. Expanding the exponentials and using the normalization of \eqref{eqn:CYnormalizations}, the central charge reduces to
\begin{equation}
Z_\gamma ~ = ~ \dfrac{k}{6} t^3 - \dfrac{k}{2}U_1 t^2 + \left(\dfrac{k}{2} U^2_1  + \dfrac{c_2 \cdot P}{24} + U_2\right) t - \left( U^3_1\dfrac{k}{6} + U_1U_2 + \dfrac{\mathcal{B}U_1}{24} + U_3 \right) \, .
\end{equation}
Using \eqref{eqn:centralcharge}, this allows for an identification of the corresponding charge vector of a single D6 brane as
\begin{align}\label{eqn:chargevector}
\gamma ~ &= ~ \left(p^0, p, \tilde{q}, \tilde{q}_0\right) \nonumber \\
&= ~ \left(1,U_1,-\dfrac{k}{2} U^2_1 - \dfrac{c_2 \cdot P}{24} - U_2,\dfrac{k}{6}U^3_1 + \dfrac{c_2 \cdot P}{24}U_1 + U_1 U_2 + U_3\right) \nonumber \\
&= ~ \left(1,U,-\dfrac{k}{2} U^2 - \dfrac{c_2 \cdot P}{24},\dfrac{k}{6}U^3 + \dfrac{c_2 \cdot P}{24}U\right) \, .
\end{align}
where in the last line, I restrict to an Abelian gauge bundle and label the only available integer class $U_1$ by $U$. This turns out to be sufficient for the polar sector of interest. Now, solving the attractor equations for a large black hole with the above charges results in a Bekenstein-Hawking entropy $S = \pi \left|Z_\gamma (t_{attractor})\right|^2$\textemdash where $t_{attractor}$ is the attractor value of the modulus determined in terms of the charges\textemdash as follows \cite{Shmakova:1996nz}
\begin{equation}
 S ~ = ~ \pi \sqrt{\mathcal{D}(1, p, \tilde{q}, \tilde{q}_0)},
\end{equation}
where $\mathcal{D}(1, p, \tilde{q}, \tilde{q}_0)$ is the discriminant function given in terms of the charges as
\begin{equation}\label{eqn:sugradiscriminant}
 \mathcal{D}(1, p, \tilde{q}, \tilde{q}_0) = \dfrac{k^2}{9} \left[ 3\dfrac{(\tilde{q} p)^2}{k^2} - 18\dfrac{\tilde{q}_0\tilde{q} p}{k^2} - 9\dfrac{\tilde{q}^2_0}{k^2} - 6\dfrac{p^3\tilde{q}_0}{k} + 8\dfrac{\tilde{q}^3}{k^3} \right] 
\end{equation}
for a single center solution. For a multi-center configuration, however, the discriminant is given by \eqref{eqn:multiregularity}, where the argument of the discriminant is chosen to be the `regularity vector' appearing in the attractor equations
\begin{equation}\label{eqn:multidiscriminant}
\mathcal{D} ~ = ~ \mathcal{D}\left(\beta + \sum_{i=1}^{n} \dfrac{\alpha_i}{\left|\vec{r} - \vec{r}_i\right|}\right) \, .
\end{equation}
Positivity of the discriminant on the `regularity vector' ensures regularity of the multi-center configuration. 

\subsection{Some generalities}
It has long been argued that a $D4$ brane splits into a bound state of a $D6$ brane and an anti-$D6$ brane \cite{Denef:2007vg}. In an M-Theory setting of the case at hand, it has proved to be very difficult to write down elliptic genera for the MSW CFTs with multiple M5 branes. Therefore, in what follows, I will consider only those with a single M5 brane. This means a unit $D4$ brane charge in the charge vector. Indeed, from the charge vector \eqref{eqn:chargevector}, considering a $D6$ brane with one unit flux and a $\bar{D6}$ with no flux yields a $D4$ brane charge vector with induced lower dimensional fluxes:
\begin{align}\label{eqn:pureD4chargevector}
\alpha_{D6} ~ &= ~ \left(1,1,-\dfrac{k}{2} - \dfrac{c_2 \cdot P}{24},\dfrac{k}{6} + \dfrac{c_2 \cdot P}{24}\right) \quad \text{and} \quad \alpha_{\bar D6} ~ = ~ \left(-1,0, \dfrac{c_2 \cdot P}{24},0\right) \quad \text{give} \nonumber \\
\alpha_{D4} ~ &= ~ \alpha_{D6} + \alpha_{\bar D6} ~ = ~ \left(0,1,-\dfrac{k}{2},\dfrac{k}{6} + \dfrac{c_2 \cdot P}{24}\right) \, .
\end{align}
Of course, as a consistency check, this must match with the appropriate induced fluxes on the $D4$ brane that cancel the Freed-Witten anomaly; this is indeed satisfied. For example, specifying to the quintic threefold, which has $k = 5$ and $c_2 \cdot P = 50$, this charge vector produces the correct fluxes known from \cite{Gaiotto:2006wm}. One can now compute the interaction degrees of freedom between these two centers and check if it matches with what one expects from modularity. Before that however, consider the $\Theta_\gamma$ decomposition of the partition function again:
\begin{equation}
\mathcal{Z}\left(q,\bar{q},y\right) ~ = ~ \sum_{\gamma = 0}^n \mathcal{Z}_\gamma\left(q\right) \Theta_\gamma \left(q,\bar{q},y\right) \, .
\end{equation}
One property of this decomposition is that $\mathcal{Z}_\gamma = \mathcal{Z}_{\delta}$ for all $\gamma = -\delta$ modulo a pull-back of the second integer cohomology onto the $D4$ brane. For the quintic for instance, $n=4$ and $\mathcal{Z}_1 = \mathcal{Z}_4$, $\mathcal{Z}_2 = \mathcal{Z}_3$. Now, the pure $D4$ brane degeneracy appears as the first (or most polar) term in the $q$-expansion of $\mathcal{Z}_0$. 
\paragraph{Symmetric product orbifolds and adding $D0$ charges} To go to the next term in the expansion, one simply adds $D0$ charge. Thinking geometrically, the $D0$ brane has a moduli space of the entire threefold in consideration and demanding a bound-state with the $D4$ reduces the moduli space of a the latter; the combined moduli space yields the correct degeneracy \cite{Gaiotto:2006wm}. Adding more and more $D0$ charges results in symmetric product orbifolds of the moduli space of the $D0$ particles, namely the threefold. Owing to configurations with coinciding branes, one runs into singularities on the moduli space that need to be resolved. As was pointed out in \cite{Manschot:2010qz}, the rational refined indices overcome these subtleties of moduli space singularites. From a multi-center configuration perspective, adding $D0$ charges implies an increase in the number of centers in a configuration. And there are exponentially many of them; the number growing with the number of partitions of the $D0$ charge to be added. Nevertheless, the low-lying spectrum can still be handled. And considering all configurations satisfying regularity, an addition of $D0$ charges takes us towards the non-polar sector of $\mathcal{Z}_0$. I will work out explicit examples in the next subsection to show that counting degrees of freedom associated to all regular configurations produces the correct polar terms.

\paragraph{Rational curves and adding $D2$ charges} In order to move `vertically', so to speak, into the degeneracies in $\mathcal{Z}_1$, one adds $D2$ charges. Thinking geometrically again, adding $D2$ charges is equivalent to demanding that the D4 brane passing through rational curves. So, one computes the moduli space associated to degree `$\beta$' rational curves in conjunction with a demand that the $D4$ brane intersect them. $D2$ fluxes, however, induce $D0$ charges and the amount of induced charge had to be computed using techniques of algebraic geometry. Even in attempts to obtain the elliptic genera from supergravity split-attractor flows \cite{Collinucci:2008ht}, the amount of induced charge was needed as an input from geometry to identify the appropriate flows that contribute to the index. Notwithstanding this input, consider the most polar\footnote{It may be worth pointing out that not all partition functions necessarily have a polar term in the $q$-expansion of $\mathcal{Z}_1$.} term in the $q$-expansion of $\mathcal{Z}_1$, say $q^{-y}$. Writing this term as $q^{-x}q^z$, such that $-x+z = -y$ with $q^{-x}$ being the most polar term in $\mathcal{Z}_0$, it turns out to be sufficient to consider added $D0$ charge that corresponds to the positive integer part of $z$.\footnote{$z$ is positive in all the examples under consideration.} In the several examples under consideration, it is sufficient to consider rational curves of degree $1$ and all polar terms of such kind have $z>1$; I leave the cases with higher degree rational curves for future work. Positivity of $(z-1)$ has a geometric interpretation: it is that rational curves come with non-trivial moduli spaces only upon an induction of $D0$ charges. In fact, in the theory of Donaldson-Thomas invariants\textemdash where a Witten index enumerates invariants $N_{DT}\left(\beta,n\right)$ associated to a D2 brane wrapping a curve in the homology class $\beta$ that intersects a collection of points ascribed to $D0$ branes\textemdash there are no topological invariants associated to $N_{DT}\left(1,0\right)$ when $z>1$. That the index is correctly reproduced by looking at multi-center configurations with added $D0$ charges as I prescribe may be interpreted as supergravity's way of telling us that $N_{DT}\left(1,0\right) = 0$ whenever $z>1$. \\

In view of the previous discussion on adding $D0$ charges, it is tempting to guess that adding $D2$ charges must involve adding additional centers to charge configurations. Interestingly, a simple argument shows that a generic $D2-D0$ charge vector never binds to a $D6$ center. Consider generic $D6$ and $D2-D0$ charge vectors as follows
\begin{equation}
 \gamma_1 = \left( 1, p, -\dfrac{p^2}{2}k - \dfrac{25}{12}, \dfrac{p^3}{6}k + \dfrac{25}{12}p \right) \quad \text{and} \quad \gamma_2 = \left( 0, 0, q, q_0 \right).
\end{equation}
Their symplectic product is given by 
\begin{equation}
 \gamma_{12} = - \left( q_0 + p q \right) \, .
\end{equation}
Using the fact that the phase factor associated to them $e^{-i \phi}$ is given by
\begin{equation}
 e^{-i \phi} \sim \dfrac{\mathrm{Z}_{(\gamma_1 + \gamma_2 = \gamma)}}{\left|\mathrm{Z}_{\gamma}\right|} \, ,
\end{equation}
we have that
\begin{align}
\mathrm{Im}\left( e^{-i\phi}\mathrm{Z}_{\gamma} \right) &\sim \mathrm{Im}\left( \mathrm{Z}_{\gamma_1}\bar{\mathrm{Z}}_{\gamma_2} \right) \nonumber \\
&\sim \left( q_0 + p q \right)\mathrm{J}^3.
\end{align}
where the second line is true up to some numerical factors and only holds in the large volume limit $\mathrm{J} \gg 0$ for a threefold with positive triple-intersection $k > 0$. Wherever it needs specification, I make an arbitrary choice for the vacuum value of the modulus $t$ at infinity to be $t = 0 + 3 i$; this satisfies the large volume condition $\mathrm{J} \gg B$. Since the FI constants now have the opposite sign of the symplectic product of the corresponding charges, the integrability equations for the bound state implies that $r_{12} < 0$, which violates regularity. This implies that a bound state of $D6$ with a generic $D2-D0$ charge never occurs! One might imagine that a three center bound-state of a generic $D2-D0$ center with $D6-\bar{D6}$ might still be possible. Although it is hard to prove in full generality, one might take the previous argument as an indication that such three-center bound states generically violate regularity. In the next subsection, I explicitly show that this is true in several examples.

\subsection{Explicit elliptic genera for some Calabi-Yau threefolds}
\subsubsection{The quintic in $\mathbb{P}^4$}
The quintic threefold is defined by a degree 5 polynomial in $\mathbb{P}^4$. The topological invariants associated to the quintic are:
$\chi (X_5) = -200$, $k=5$ and $c_2 \cdot P = 50$. Its modified elliptic genus is given by 
\begin{align}
\mathcal{Z}_{X_5}\left(q,\bar{q},y\right) ~ &= ~ \sum_{\gamma = 0}^4 \mathcal{Z}_\gamma\left(q\right) \Theta^{(5)}_\gamma \left(q,\bar{q},y\right) \qquad \text{and} \nonumber \\ 
\Theta^{(m)}_k \left(q,\bar{q},y\right) ~ &= ~ \sum_{n \in \mathbb{Z} + \frac{1}{2} + \frac{k}{m}}  \left(-1\right)^{mn} q^{\frac{m}{2} n^2} y^{m n} \, 
\end{align}
where
\begin{align}
\mathcal{Z}_0 (q) ~ &= ~ q^{-\frac{55}{24}} \left(5 - 800 q + 58500 q^2 + \text{non-polar terms}\right) \nonumber \\
\mathcal{Z}_1(q) ~ &= ~ Z_4(q) ~ = ~ q^{-\frac{83}{120}} \left(8625 + \text{non-polar terms}\right) \nonumber \\
\mathcal{Z}_2(q) ~ &= ~ Z_3(q) ~ = ~ \text{non-polar terms} \, .
\end{align}

\paragraph{Pure $D4$ brane}
The charge vector associated to a Pure $D4$ brane for this compactification can be written from \eqref{eqn:pureD4chargevector} with the topological data of the quintic
\begin{align}\label{eqn:PureD4quintic}
\gamma_1 ~ \coloneqq ~ \alpha_{D6} ~ &= ~ \left(1,1,-\dfrac{55}{12},\dfrac{35}{12}\right) \quad \text{and} \quad \gamma_2 ~ \coloneqq ~ \alpha_{\bar D6} ~ = ~ \left(-1,0, \dfrac{25}{12},0\right) \quad \text{give} \nonumber \\
\gamma ~ \coloneqq ~ \alpha_{D4} ~ &= ~ \alpha_{D6} + \alpha_{\bar D6} ~ = ~ \left(0,1,-\dfrac{5}{2},\dfrac{35}{12}\right) \, .
\end{align}
Computing the discriminant associated to this vector via \eqref{eqn:sugradiscriminant}, one finds
\begin{equation}\label{eqn:D4discriminantvalue}
 \mathcal{D}_{D4} = -\dfrac{275}{36} \, ,
\end{equation}
which yields an imaginary single-center entropy. This renders this solution un-physical\footnote{In fact, computing the discriminant associated to the $D6$ center also yields a negative value: $-3125/1944$. I expect that the ignored instanton corrections to the prepotential lift this sickness; working with this hypothesis, I merely shift the definition of a `zero discriminant' from $\mathcal{D}_\gamma = 0$ to that of the $D6$ brane. Aside from this subtlety, the instanton corrections play no other role in the analysis.}. In order to compute the interaction degrees of freedom, the two-center integrability equations
\begin{equation}\label{eqn:D4Denef}
 \dfrac{\gamma_{12}}{z_{12}} = c_1,
\end{equation}
where $z_{12} \in \mathbb{R}$, need to be solved. The required constants, to solve this equation, are tabulated below in \autoref{tab:pureD4data}.
\begin{table}[!h]
\begin{center}
\begin{tabular}{|c|c|c|c|c|c|}
  \hline
  $\gamma_{12}$ & $\mathrm{Z}_1$ & $\mathrm{Z}_2$ & $\alpha = \arg[Z_{\gamma}]$ & $\mathcal{D}_{\gamma}$ & $c_1$\\
  \hline \hline
  -5 & $\left( \dfrac{47}{72} + i\dfrac{7}{24} \right)\sqrt{5}$ & $-i\dfrac{13\sqrt{5}}{24}$ & $\tan^{-1}{\left(\dfrac{18}{47}\right)}$ & $-\dfrac{275}{36}$ & $-\dfrac{611\sqrt{\dfrac{5}{2533}}}{12}$\\ 
  \hline 
\end{tabular}  
    \caption{Relevant constants for the Pure $D4$ brane.}
    \label{tab:pureD4data}
\end{center}
\end{table}
This results in the following solution
\begin{equation}
 z_{12} = \dfrac{611\sqrt{12665}}{12} \, .
\end{equation}
Since it is only the relative distance between the centers that is important, I fix $z_1$ to be at the origin. The above solution then implies that $z_2$ is at a distance of $\pm z_{12}$ from the origin on the axis on which the centers are localized. This leaves us with two possible configurations, namely: $12$ and $21$, where $z_1 < z_2$ and $z_2 < z_1$ respectively.\footnote{In the configuration $12$, $z_2 = + z_{12}$ and in the configuration $21$, $z_2 = - z_{12}$} For consistency, the discriminant associated to the two-center configuration must be positive. This requires the knowledge of $\beta$\textemdash with an arbitrary choice of the value for the modulus at infinity to be $t = 0 + 3i$ as mentioned before\textemdash \\
\begin{equation}
 \beta = \left( \dfrac{6}{\sqrt{12665}}, \dfrac{53}{\sqrt{12665}}, - \dfrac{611}{12}\sqrt{\dfrac{5}{2533}}, - \dfrac{76}{\sqrt{12665}} \right),
\end{equation}
Plugging this into the regularity vector, I find that both configurations $\mathcal{D}_{12}$ and $\mathcal{D}_{21}$ are regular everywhere\footnote{For simplicity, I check for positivity of the corresponding regularity vector only along the axis of localization.} outside the centers; infinities at the location of the centers is expected. Given all the configurations that contribute, using the formula in \eqref{eqn:gref}
\begin{equation}
 g_{\text{ref}}(\{ \gamma_i \}, y) = \dfrac{(-1)^{\sum_{i < j} \gamma_{ij}+n-1}}{\left(y-y^{-1}\right)^{n-1}} \displaystyle\sum_{p} s(p) ~ y^{\sum_{i < j}\gamma_{ij}sign[z_j -z_i]} \, ,
\end{equation}
the Poincar\'{e} polynomial associated to the interaction degrees of freedom of the Pure $D4$ brane realized as a bound state of the $D6$ and $\bar{D6}$ is
\begin{align}\label{eqn:pureD4gref}
 g_{\text{ref}}( \gamma_1, \gamma_2, y) &= \dfrac{(-1)^{-5 + 2 - 1}}{\left(y-y^{-1}\right)} \left(y^5-y^{-5}\right) \nonumber \\
 &= \left( y^{-4} + y^{-2} + 1 + y^2 + y^4 \right) \, ,
\end{align}
where the sign $s(12)$ was computed to be + 1 from the Hessian of the superpotential in \eqref{eqn:superpot}. Specializing to $y \to (-1)$ results in $g_{\mathrm{ref}} = 5$. Substituting this into \eqref{eqn:finalindex} with the implicit understanding that a $D6$ and a $\bar{D6}$ have refined indices of $1$ each\footnote{The structure sheaves have a unit degeneracy.} yields a final index of 
\begin{equation}
 \bar \Omega(\gamma_1, \gamma_2;t) ~ = ~ g_{\mathrm{ref}}~\bar{\Omega}^S_{D6}~\bar{\Omega}^S_{\bar{D6}} ~ = ~ 5 \times 1 \times 1 ~ = ~ 5.
\end{equation}
This matches the prediction from the string regime and modularity; the most polar term in $\mathcal{Z}_0$ is the pure $D4$ brane. The final index being exactly the same as the norm of the symplectic inner product of the two charge vectors is not a mere coincidence. This is a generic feature of two center solutions to the integrability equations.

 \paragraph{$D4$-$D0$ bound state}
As advertised in the previous subsection, the next polar term in $\mathcal{Z}_0$ may be achieved by adding a D0 brane center. Three-center solutions of $D0$ branes bound to $D6$ and $\bar{D6}$ centers have been extensively studied in \cite{Manschot:2011xc}. While the $D6$ centers considered there were both with $D4$ fluxes turned on, the analysis is largely similar. It has also been previously noted that $D_{p-6}$ branes bound to $D_p$ branes energetically prefer to stay ejected from them as opposed to dissolving as fluxes as preferred by $D_{p-2}$ and $D_{p-4}$ branes. This is consistent with the picture in \cite{Manschot:2011xc} that adding a $D0$ charge necessarily implies an addition of a new $D0$ center with charge vector $\alpha_{D0} = \left(0,0,0,\pm 1\right)$. The correct sign may be fixed by noting that adding a positive $D0$ flux on the Pure $D4$ reduces the entropy via a reduction in $\mathcal{D}$. Therefore, in these conventions, a $D4$ brane binds to an anti-$D0$ brane. Therefore, the three-problem of interest now has a third center $\gamma_3 \coloneqq \left(0,0,0,-1\right)$ in addition to the two centers that generated a pure $D4$ brane. These result in a total charge vector given by
\begin{equation}
 \gamma ~ \coloneqq ~ \alpha_{D4-D0} = \left( 0, 1, -\dfrac{5}{2}, \dfrac{23}{12} \right).
\end{equation}
The corresponding integrability equations take the form
\begin{equation}
 \dfrac{\gamma_{12}}{z_{12}} + \dfrac{\gamma_{13}}{z_{13}} = c_1,
\end{equation}
\begin{equation}
 \dfrac{\gamma_{23}}{z_{23}} + \dfrac{\gamma_{21}}{z_{21}} = c_2,
\end{equation}
where $\gamma_{12} = -\gamma_{21}$\footnote{The symplectic product of any two charge vectors is antisymmetric.}. I tabulate the relevant data required to solve these equations, in tables \autoref{tab:D4-D0data} and \autoref{tab:D4-D0datacontd.}.
\begin{table}[!h]
\begin{center}
\begin{tabular}{|c|c|c|c|c|c|c|}
  \hline
  $\gamma_{12}$ & $\gamma_{13}$ & $\gamma_{23}$ & $\mathrm{Z}_1$ & $\mathrm{Z}_2$ & $\mathrm{Z}_3$ & $\alpha = \arg[Z_{\gamma}]$\\
  \hline \hline
  -5 & 1 & -1 & $\left( \dfrac{47}{72} + i\dfrac{7}{24} \right)\sqrt{5}$ & $-i\dfrac{13\sqrt{5}}{24}$ & $\dfrac{1}{6\sqrt{5}}$ & $\tan^{-1}{\left(\dfrac{90}{247}\right)}$ \\ 
  \hline 
\end{tabular}
  \caption{Relevant data for the $D4-D0$ state (Part a).}
  \label{tab:D4-D0data}
\end{center}
\end{table}
\begin{table}[!h]
\begin{center}
\begin{tabular}{|c|c|c|}
  \hline
  $\mathcal{D}_{\gamma}$ & $c_1$ & $c_2$\\
  \hline \hline
  $-\dfrac{155}{36}$ & $-\dfrac{3139}{12}\sqrt{\dfrac{5}{69109}}$ & $\dfrac{3211}{12}\sqrt{\dfrac{5}{69109}}$\\ 
  \hline 
\end{tabular}
  \caption{Relevant data for the $D4$-$D0$ state (Part b).}
  \label{tab:D4-D0datacontd.}
\end{center}
\end{table}
Starting far out in the moduli space at $t = 0 + 3i$ again, the corresponding vector for $\beta$ is
\begin{equation}
 \beta = \left( -6\sqrt{\dfrac{5}{69109}}, \dfrac{247}{\sqrt{345545}}, -\dfrac{3211}{12}\sqrt{\dfrac{5}{69109}}, 2\left( -\dfrac{295}{4}\sqrt{\dfrac{5}{69109}} + \dfrac{247}{4\sqrt{345545}} \right) \right).
\end{equation}
Solving the Denef equations and writing down those solutions that satisfy the discriminant positivity condition (i.e, $\mathcal{D} > \mathcal{D}_{D6}$) I find \autoref{tab:D4-D0solns}.
\begin{table}[!h]
\begin{center}
\begin{tabular}{|c||c|c|c|c|}
  \hline
  Configuration & $z_1$ & $z_2$ & $z_3$ & $s(p)$\\
  \hline \hline
  $231$ & $0$ & $-1.33278$ & $-0.65506$ & -1\\
  \hline
  $312$ & $0$ & $2.07521$ & $-5.42298$ & 1\\
  \hline
  $132$ & $0$ & $1.33278$ & $0.65506$ & -1\\
  \hline
  $213$ & $0$ & $-2.07521$ & $5.42298$ & 1\\
  \hline 
\end{tabular}
  \caption{Configurations contributing to the $D4$-$D0$ state.}
  \label{tab:D4-D0solns}
\end{center}
\end{table}
Gathering all the computations, I now compute the interaction degrees of freedom for this three center bound state
\begin{align}\label{eqn:D4-D0gref}
 g_{\text{ref}}( \gamma_1, \gamma_2, \gamma_3, y) &= \dfrac{(-1)^{-5 + 1 - 1}}{\left(y-y^{-1}\right)^2} \left(y^5-y^3-y^{-3}+y^{-5}\right) \nonumber \\
 &= -\left( y^{-3} + y^{-1} + y^1 + y^3 \right) \, .
\end{align}
Specializing to $y \to (-1)$ results in $g_{\text{ref}} = 4$. Substituting this into \eqref{eqn:finalindex} and using the fact that the single center refined index for a $D0$ is $\chi(CY_3) = -200$ yields a final index of

\begin{align}\label{finalD4-D0index}
 \bar \Omega(\gamma_1, \gamma_2, \gamma_3; t) &= g_{\text{ref}}~\bar{\Omega}^S_{D6}~\bar{\Omega}^S_{\bar{D6}}~\bar{\Omega}^S_{D0} \nonumber \\
 &= 4 \times 1 \times 1 \times (-200) \nonumber \\
 &= -800.
\end{align}
This too is in perfect agreement with the partition function.

\paragraph{$D4$-$D0$-$D0$ bound states}
There are two possibilities for the next polar state.
\begin{itemize}
 \item A three center scenario, similar to the $D4$-$D0$ case\footnote{This is an example of a scenario where $\mathcal{D} > \mathcal{D}_{D6}$ and yet it corresponds to a purely multi-center solution.}, but with the third center carrying twice the unit $D0$ charge $\alpha_{2D0} = \left( 0, 0, 0, -2 \right)$. The total charge vector is now
\begin{equation}
 \gamma = \left( 0, 1, -\dfrac{5}{2}, \dfrac{11}{12} \right).
\end{equation}
The corresponding integrability equations take the form
\begin{equation}
 \dfrac{\gamma_{12}}{z_{12}} + \dfrac{\gamma_{13}}{z_{13}} = c_1,
\end{equation}
\begin{equation}
 \dfrac{\gamma_{23}}{z_{23}} + \dfrac{\gamma_{21}}{z_{21}} = c_2.
\end{equation}
The relevant data required to solve these equations is collected in the following tables.
\begin{table}[!h]
\begin{center}
\begin{tabular}{|c|c|c|c|c|c|c|}
  \hline
  $\gamma_{12}$ & $\gamma_{13}$ & $\gamma_{23}$ & $\mathrm{Z}_1$ & $\mathrm{Z}_2$ & $\mathrm{Z}_3$ & $\alpha = \arg[Z_{\gamma}]$\\
  \hline \hline
  -5 & 2 & -2 & $\left( \dfrac{47}{72} + i\dfrac{7}{24} \right)\sqrt{5}$ & $-i\dfrac{13\sqrt{5}}{24}$ & $\dfrac{1}{3\sqrt{5}}$ & $\tan^{-1}{\left(\dfrac{90}{259}\right)}$ \\ 
  \hline 
\end{tabular}
  \label{tab:D4-2D0data}
  \caption{Relevant data for the $D4$-$2D0$ state (Part a).}
\end{center}
\end{table}

\begin{table}[!h]
\begin{center}
\begin{tabular}{|c|c|c|}
  \hline
  $\mathcal{D}_{\gamma}$ & $c_1$ & $c_2$\\
  \hline \hline
  $-\dfrac{35}{36}$ & $-\dfrac{3223}{12}\sqrt{\dfrac{5}{75181}}$ & $\dfrac{3367}{12}\sqrt{\dfrac{5}{75181}}$\\ 
  \hline 
\end{tabular}
  \label{tab:D4-2D0datacontd.}
  \caption{Relevant data for the $D4$-$2D0$ state (Part b).}
\end{center}
\end{table}
The corresponding vector for $\beta$ is
\begin{equation}
 \beta = \left( -6\sqrt{\dfrac{5}{75181}}, \dfrac{259}{\sqrt{375905}}, -\dfrac{3367}{12}\sqrt{\dfrac{5}{75181}}, 2\left( -\dfrac{295}{4}\sqrt{\dfrac{5}{75181}} + \dfrac{259}{4\sqrt{375905}} \right) \right).
\end{equation}
Solving the integrability equations and writing down those solutions that satisfy the discriminant positivity condition, I find the values in Table \ref{tab:D4-2D0solns}.
\begin{table}[!h]
\begin{center}
\begin{tabular}{|c||c|c|c|c|}
  \hline
  Configuration & $z_1$ & $z_2$ & $z_3$ & $s(p)$\\
  \hline \hline
  $231$ & $0$ & $-0.446522$ & $-0.222042$ & -1\\
  \hline
  $312$ & $0$ & $1.95346$ & $-5.41678$ & 1\\
  \hline
  $132$ & $0$ & $0.446522$ & $0.222042$ & -1\\
  \hline
  $213$ & $0$ & $-1.95346$ & $5.41678$ & 1\\
  \hline 
\end{tabular}
  \caption{Configurations contributing to the $D4$-$2D0$ state.}
  \label{tab:D4-2D0solns}
\end{center}
\end{table}
The associated Poincar\'{e} polynomial is now
\begin{align}\label{eqn:D4-2D0gref}
 g_{\text{ref}}( \gamma_1, \gamma_2, \gamma_3, y) &= \dfrac{(-1)^{-5 + 2 - 2}}{\left(y-y^{-1}\right)^2} \left(y^5-y^1-y^{-1}+y^{-5}\right) \nonumber \\
 &= -\left( y^{-3} + 2y^{-1} + 2y^1 + y^3 \right) \, .
\end{align}
Therefore $g_{\text{ref}} = 6$ while the refined index for the 2$D0$ center is given by 
\begin{equation}
 \bar{\Omega}^S_{2D0} = \chi(CY_3) + \dfrac{\chi(CY_3)}{4} = -250.
\end{equation}
This yields a final index of
\begin{align}\label{finalD4-2D0index}
 \bar \Omega(\gamma_1, \gamma_2, \gamma_3; t) &= g_{\text{ref}}~\bar{\Omega}^S_{D6}~\bar{\Omega}^S_{\bar{D6}}~\bar{\Omega}^S_{2D0} \nonumber \\
 &= 6 \times 1 \times 1 \times (-250) \nonumber \\
 &= -1500.
\end{align}

 \item A four center scenario with two explicit $D0$ centers:\\
 
The contributing centers are the previous $D6$ and $\bar{D6}$ centers with two explicit unit charge $D0$ charge vectors. The total charge vector is clearly the same as before. A detailed computation is no more illuminating to present here; the resulting $g_{\text{ref}}$ in this scenario is half that of the previous case, owing to the halving of the symplectic products. The degeneracy for this four center $D6$-$\bar{D6}$-$D0$-$D0$ solution is $g_{\text{ref}} = 3$. This yields a final index of
\begin{align}\label{finalD4-D0-D0index}
 \bar{\Omega}(\gamma_1, \gamma_2, \gamma_3, \gamma_4; t) &= \dfrac{g_{\text{ref}}}{2}~\bar{\Omega}^S_{D6}~\bar{\Omega}^S_{\bar{D6}}~\bar{\Omega}^S_{D0}~\bar{\Omega}^S_{D0} \nonumber \\
 &= \dfrac{3}{2} \times 1 \times (-200) \times (-200) \nonumber \\
 &= 60000,
\end{align}
where the factor of half comes from the automorphism arising from the two identical $D0$ centers. This results in a total contribution of -1500 + 60000 = 58500, towards this state. All these numbers are clearly consistent with the modular prediction for $\mathcal{Z}_0$.
\end{itemize}
Once one has identified the appropriate centers that are of interest, the authors of \cite{Manschot:2013dua} have developed a Mathematica code for the computation of the Poincare polynomials. The code is attached to their paper.

\begin{figure}[h!]
  \includegraphics[width=\linewidth]{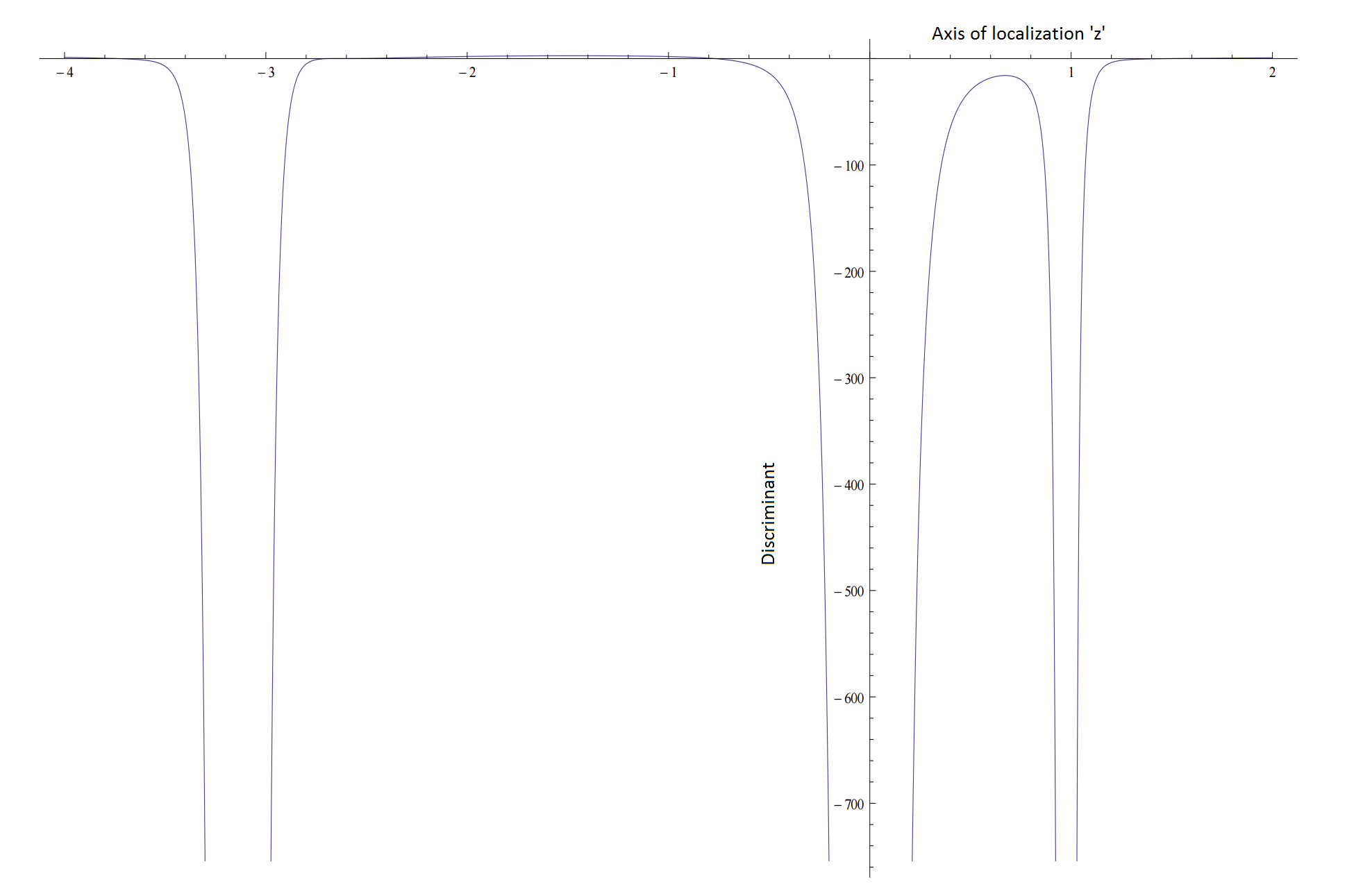}
  \caption{Discriminant function of a $D6-\bar{D6}-D2_{D0}$ three-center configuration with charges $D6$ and $\bar{D6}$ as in \eqref{eqn:PureD4quintic} and $D2_{D0}$ charge vector $(0,0,-1,2)$ plotted against the axis of localization of the centers.}
  \label{fig:D2plot}
\end{figure}

\paragraph{$D6$-$\bar{D6}$-$D2_{D0}$ bound states}
To move into the polar sector of $\mathcal{Z}_1$, now, one must add $D2$ charges. Naively, this added charge may merely be an increase in the $D2$ component of either of the $D6$ brane charges or act as an additional third center, with possible additional induced $D0$ centers. The split attractor flow allows for such flows into many channels \cite{Denef:2007vg}. In fact, in the large volume limit\textemdash the one I stick to in this article\textemdash one can even compute the index across the wall of marginal stability along the flow \cite{Manschot:2010xp}; for an end point with three centers\textemdash which I will think of being the two $D6$ centers along with a generic $D2_{D0}$ center $\left(0,0,q,q_0\right)$\textemdash is given by \cite{Manschot:2010xp}
\begin{align}
 \Omega((12)3;t) ~  &= ~ \dfrac{1}{4} (-1)^{\tiny \gamma_{12} + \gamma_{31} + \gamma_{23}} \left. \gamma_{\tiny (1+2)3} \cdot \gamma_{\tiny 12} \cdot \Omega(\gamma_1) \cdot \Omega(\gamma_2) \cdot     \Omega(\gamma_3) \right. \nonumber \\
 & \quad \left. \quad \left(\underbrace{\mathrm{sgn}\left[\mathrm{Im}[Z(\gamma_1+\gamma_2,t)\bar{Z}(\gamma_3,t)]\right]}_{a} + \underbrace{\mathrm{sgn}[\gamma_{\tiny (1+2)3}]}_{b}\right) \right. \nonumber \\
 & \quad \left. \qquad \left(\underbrace{\mathrm{sgn}\left[\mathrm{Im}[Z(\gamma_1,t_1)\bar{Z}(\gamma_2,t_1)]\right]}_{c} + \underbrace{\mathrm{sgn}[\gamma_{\tiny 12}]}_{d}\right) \right..
\end{align}
Specifying the quintic data, I find
\begin{align}
 \mathrm{sgn}\left[a\right] ~ &= ~ \mathrm{sgn}\left[q_0 - \dfrac{235}{12}\right] \, ; \quad \mathrm{sgn}\left[b\right] ~ &&= ~ -\mathrm{sgn}\left[5 + q_0\right] \nonumber \\
 \mathrm{sgn}\left[d\right] ~ &= ~ -\mathrm{sgn}\left[q + q_0\right] \, ; \quad ~~  \mathrm{sgn}\left[c\right] ~ &&= ~ \mathrm{sgn}\left[\dfrac{-72q^2_0 - 360q_0 + 408qq_0 + 5785q}{1728(5 + q_0)}\right] \, .
\end{align}
upon computing the corresponding quantities in the underbraces. For a non-vanishing index, $\mathrm{sgn}\left[a\right]$ and $\mathrm{sgn}\left[b\right]$ must have the same sign (and similarly with $\mathrm{sgn}\left[c\right]$ and $\mathrm{sgn}\left[d\right]$). $\mathrm{sgn}\left[a\right]$ and $\mathrm{sgn}\left[b\right]$ have the same sign iff $-5 < q_0 < 20$, where I use the fact that $q$ \& $q_0 \in \mathbb{Z}$. Therefore, if this condition is satisfied, $\mathrm{sgn}\left[a\right] + \mathrm{sgn}\left[b\right] = -2$. For the total contribution to the index to be positive\footnote{Considerations similar to those that will follow, rule out the case when the contribution is negative too.},
$\mathrm{sgn}\left[c\right] ~ \mathrm{and} ~ \mathrm{sgn}\left[d\right] < 0$. Since $\gamma_{12} = -q - q_0$,
\begin{equation}
\mathrm{sgn}\left[d\right] = \mathrm{sgn}\left[\gamma_{12}\right] = - \mathrm{sgn}\left[q + q_0\right].
\end{equation}
Now $\mathrm{sgn}\left[d\right] < 0$ implies $q > - q_0$. Putting all the pieces together, the allowed values for the center $\gamma_2$ such that there is a non-vanishing contribution to the index are collected in \autoref{tab:D2splitflowtable}. It is evident that in the direction of the physical $D0$ charges that bind with the $D4$, there are no non-vanishing $D2$ charges to form a three-center black hole bound state. Nevertheless, one might still investigate if $D0$ charges of the opposite sign can form the third center with non-vanishing $D2$ charges. As it turns out, none of the allowed values in \autoref{tab:D2splitflowtable} result in a positive discriminant everywhere outside the location of the centers. An example of this is shown in \autoref{fig:D2plot} where the discriminant function \eqref{eqn:multidiscriminant} associated to a three center configuration with charges $D6$ and $\bar{D6}$ as in \eqref{eqn:PureD4quintic} and a third $D2_{D0}$ center with $(0,0,-1,2)$ is plotted against the axis of localization of the centers. Owing to the negative discriminant of the Pure D4 brane (arising from the ignoring of instanton corrections to the prepotential), one expects that the discriminant is negative at the locations of the $D6$ and $\bar{D6}$ centers. However, as is evident from the plot, the discriminant dips below zero even near the third center corresponding to the $D2_{D0}$ charge vector. This charge vector has zero discriminant and therefore must not go down to negative infinity as it does in the plot. One may easily check that in fact all allowed values of the $D2_{D0}$ center listed in \autoref{tab:D2splitflowtable} violate regularity.

This rules out the possibility of having a three center bound state with non-vanishing $D2$ charges. This may be seen as a more precise vindication of the naive argument I presented in the previous subsection. \\

In so far as the modified elliptic genus is concerned, this means that a polar term with $D2$ charges can occur only as a two-center configuration where one of the centers has additional $D2$ and $D0$ fluxes. To identify which of the two centers picks up the additional lower dimensional charges, I again look for the charge vectors whose discriminant increases upon the addition of the said charges to find the configuration to be
\begin{equation}\label{eqn:nontriv.singlecenter}
 D6 \colon \quad \gamma_1 = \left( 1, 1, -\dfrac{55}{12}, \dfrac{23}{12} \right) \, \qquad \text{and} \qquad \bar{D6} \colon \quad \gamma_2 = \left( -1, 0, \dfrac{13}{12}, 0 \right). \qquad
\end{equation}
Since this is now a two center problem, $g_{ref}$ is given by the symplectic product of the charge vectors $g_{\text{ref}} = \left|\gamma_{12}\right| = 3$. Therefore, the final index is given by
\begin{align}
 \bar{\Omega}(\gamma_1, \gamma_2; t) &= g_{\text{ref}}~\bar{\Omega}^S_1~\bar{\Omega}^S_2 \nonumber \\
 &= 3 \times 2875 \times 1 \nonumber \\
 &= 8625,
\end{align}
where the factor of 2875 comes from the Donaldson-Thomas invariants associated to the $D6$ with a $D2$ flux and a point $p$.

\subsubsection{$X_6$ in $\mathbb{WP}_{(2,1,1,1,1)}$}
The sextic is a degree 6 hypersurface in $\mathbb{WP}_{(2,1,1,1,1)}$. For the purposes of this article, the topological invariants associated to the sextic I need are:
$\chi (X_6) = -204$, $k=3$, $c_2 \cdot P = 42$ and $N_{DT}(1,1)=7884$. Its modified elliptic genus is given by 
\begin{equation}
\mathcal{Z}_{X_6}\left(q,\bar{q},y\right) ~ = ~ \sum_{\gamma = 0}^2 \mathcal{Z}_\gamma\left(q\right) \Theta^{(3)}_\gamma \left(q,\bar{q},y\right) 
\end{equation}
where
\begin{align}
\mathcal{Z}_0 (q) ~ &= ~ q^{-\frac{45}{24}} \left(4 - 612 q + \text{non-polar terms}\right) \nonumber \\
\mathcal{Z}_1(q) ~ &= ~ Z_2(q) ~ = ~ q^{-\frac{5}{24}} \left(15768 + \text{non-polar terms}\right) \, .
\end{align}
The charge vector of the Pure $D4$ brane is given by
\begin{align}
\alpha_{D6} ~ &= ~ \left(1,1,-\dfrac{13}{4},\dfrac{9}{4}\right) \quad \text{and} \quad \alpha_{\bar D6} ~ = ~ \left(-1,0, \dfrac{7}{4},0\right) \quad \text{give} \nonumber \\
\alpha_{D4} ~ &= ~ \alpha_{D6} + \alpha_{\bar D6} ~ = ~ \left(0,1,-\dfrac{3}{2},\dfrac{9}{4}\right) \, ,
\end{align}
Omitting explicit detail, the associated Poincar\'{e} polynomial is
\begin{equation}
g_{\text{ref}} \left(D4,y\right) ~ = ~ -\left(y^{-3} + y^{-1} + y^1 + y^3 \right) \, .
\end{equation}
which yields the correct final index of 4. Adding a $D0$ brane, yields 
\begin{equation}
g_{\text{ref}} \left(D4-D0,y\right) ~ = ~ y^{-2} + 1 + y^2 \, 
\end{equation}
which gives a final index of $3 \times -204 = -612$. Finally, adding a $D2$ charge, the two centers are
\begin{equation}
D6 \colon \quad \gamma_1 = \left( 1, 1, -\dfrac{13}{4}, \dfrac{5}{4} \right) \, \qquad \text{and} \qquad \bar{D6} \colon \quad \gamma_2 = \left( -1, 0, \dfrac{3}{4}, 0 \right). \qquad
\end{equation}
with
\begin{equation}
g_{\text{ref}} \left(D4-D2_{D0},y\right) ~ = ~ 2 \, 
\end{equation}
yielding a final index of $2 \times 7884 = 15768$.
\subsubsection{$X_8$ in $\mathbb{WP}_{(4,1,1,1,1)}$}
The octic threefold is a degree 8 hyperplane in $\mathbb{WP}_{(4,1,1,1,1)}$ its relevant topological invariants are:
$\chi (X_8) = -296$, $k=2$, $c_2 \cdot P = 44$ and $N_{DT}(1,1)=29504$. Its modified elliptic genus is given by 
\begin{equation}
\mathcal{Z}_{X_8}\left(q,\bar{q},y\right) ~ = ~ \sum_{\gamma = 0}^4 \mathcal{Z}_\gamma\left(q\right) \Theta^{(2)}_\gamma \left(q,\bar{q},y\right)
\end{equation}
where
\begin{align}
\mathcal{Z}_0 (q) ~ &= ~ q^{-\frac{23}{12}} \left(4 - 888 q + \text{non-polar terms}\right) \nonumber \\
\mathcal{Z}_1(q) ~ &= ~ q^{-\frac{1}{6}} \left(59008 + \text{non-polar terms}\right) \, .
\end{align}
The Pure $D4$ brane is now
\begin{align}
\alpha_{D6} ~ &= ~ \left(1,1,-\dfrac{17}{6},\dfrac{13}{6}\right) \quad \text{and} \quad \alpha_{\bar D6} ~ = ~ \left(-1,0, \dfrac{11}{6},0\right) \quad \text{give} \nonumber \\
\alpha_{D4} ~ &= ~ \alpha_{D6} + \alpha_{\bar D6} ~ = ~ \left(0,1,-1,\dfrac{11}{6}\right) \, ,
\end{align}
with an associated Poincar\'{e} polynomial
\begin{equation}
g_{\text{ref}} \left(D4,y\right) ~ = ~ -\left(y^{-3} + y^{-1} + y^1 + y^3 \right) \, .
\end{equation}
and final index of $4$. Adding a $D0$ brane, yields 
\begin{equation}
g_{\text{ref}} \left(D4-D0,y\right) ~ = ~ y^{-2} + 1 + y^2 \, 
\end{equation}
which gives a final index of $3 \times -296 = -888$. Finally, adding a $D2$ charge, the two centers are
\begin{equation}
D6 \colon \quad \gamma_1 = \left( 1, 1, -\dfrac{17}{6}, \dfrac{7}{6} \right) \, \qquad \text{and} \qquad \bar{D6} \colon \quad \gamma_2 = \left( -1, 0, \dfrac{5}{6}, 0 \right). \qquad
\end{equation}
with
\begin{equation}
g_{\text{ref}} \left(D4-D2_{D0},y\right) ~ = ~ 2 \, 
\end{equation}
yielding a final index of $2 \times 29504 = 59008$.
\subsubsection{$X_{10}$ in $\mathbb{WP}_{(5,2,1,1,1)}$}
The dectic is a degree 10 hypersurface in $\mathbb{WP}_{(5,2,1,1,1)}$. The relevant topological invariants associated to the dectic are:
$\chi (X_{10}) = -288$, $k=1$ and $c_2 \cdot P = 34$. Its modified elliptic genus is given by 
\begin{align}
\mathcal{Z}_{X_5}\left(q,\bar{q},y\right) ~ &= ~ \dfrac{\eta(q)^{-35}}{576} \left[541 E_4(q)^4 + 1187 E_4(q) E_6(q)^2\right] \Theta_1(\bar{q},y) \nonumber \\  
&= ~ q^{-\frac{35}{24}} \left(3 - 576 q + \text{non-polar terms}\right) \, .
\end{align}
A Pure $D4$ brane in this example is
\begin{align}
\alpha_{D6} ~ &= ~ \left(1,1,-\dfrac{23}{12},\dfrac{19}{12}\right) \quad \text{and} \quad \alpha_{\bar D6} ~ = ~ \left(-1,0, \dfrac{17}{12},0\right) \quad \text{give} \nonumber \\
\alpha_{D4} ~ &= ~ \alpha_{D6} + \alpha_{\bar D6} ~ = ~ \left(0,1,-\dfrac{1}{2},\dfrac{19}{12}\right) \, .
\end{align}
The Poincar\'{e} polynomial is
\begin{equation}
g_{\text{ref}} \left(D4,y\right) ~ = ~ y^{-2} + 1 + y^2 \, .
\end{equation}
And the corresponding index is $3$. Adding a $D0$ brane, yields 
\begin{equation}
g_{\text{ref}} \left(D4-D0,y\right) ~ = ~ - y^{-1} - y^1 \, 
\end{equation}
which gives a final index of $2 \times -288 = -576$. 
Clearly, all results exactly build the polar terms under consideration in the examples.

\section{Discussion}\label{sec:discussion}
In this article, I have identified all multi-center configurations (whose total charge vectors violate the naive single-center cosmic censorship bound) that build the polar sector of several elliptic genera of Calabi-Yau threefolds with maximal holonomy. It is natural to expect that once one moves into the non-polar sector of the theory, when total charge vectors no longer violate the cosmic-censorship bound, single center black holes begin to contribute. Exactly what states these constitute is not fully known. Several interesting suggestions have been made \cite{Bena:2012hf,Lee:2013yka,Lee:2012sc,VanHerck:2009ww} in the literature. Nevertheless, large charge single-center black hole entropy has not been easy to understand concretely, with these suggestions. \\

With the prescription I have proposed in this article, one may now seek to push into the non-polar sector of the elliptic genus to understand single-center black hole entropy. Naively, the approach from split-flows proposed in \cite{Collinucci:2008ht,VanHerck:2009ww} might have been a good starting point to push deep into the non-polar sector. As one increases charge, there is an expectation that an increasing number of multi-center configurations must contribute to the index. For instance, moving on from a $D4$-$D0$ charge vector to a $D4$-$2D0$ charge vector, one expects two different contributions: one from a three center $D6$-$\bar{D6}$-$2D0$ solution and another from a four center $D6$-$\bar{D6}$-$D0$-$D0$ configuration. The authors of \cite{Collinucci:2008ht,VanHerck:2009ww}, however, argue for only a single flow. On the contrary, I show explicitly in \eqref{finalD4-2D0index} and \eqref{finalD4-D0-D0index} that both the expected configurations do indeed contribute to produce the correct polar term. In extension, enumeration of all multi-center configurations can systematically be done with the approach I present in this article. It may be noted that a more general prescription might be needed to incorporate higher degree rational curves to include more $D2$ charges with appropriately induced $D0$ charges. Nevertheless, it is my hope that this enables for a better understanding of the non-polar sector. \\

Although I have refrained from stressing on them, there are several aspects of purely mathematical interest that are very closely related to the study of BPS states mentioned in this article. The elliptic genera studied in this article encode topological invariants of the moduli space of the derived category of coherent sheaves on a Calabi-Yau threefold. Some questions in this field related to this article are: What is the generating function for the Euler numbers of the moduli space of stable sheaves (seen as objects in the derived category of coherent sheaves where stability is usually thought to be Bridgeland stability as in the Kontsevich-Soibelman setup) on a smooth three-dimensional quasi-projective variety? What is the generating function for the Betti numbers for the same? G\"{o}ttsche has answered both these questions for sky-scraper sheaves on smooth two-dimensional quasi-projective varieties \cite{Göttsche1990}. In the mid-nineties, Cheah \cite{Cheah} managed to write a generating function (the McMahon function) for the Euler numbers of the moduli space of stable sky-scraper sheaves on smooth three dimensional quasi-projective varieties. A refinement of Cheah's result in the spirit of G\"{o}ttsche is expected to be related to single-center black hole entropies. While hoping for a general result might be far fetched from the explicit multi-center prescription presented in this article, I speculate that it may well prove to be very helpful in conjuring up and testing conjectures \cite{Bena:2012hf,Lee:2013yka} in this regard. \\ 

Another branch of mathematical interest that is closely related is the study of Poincar\'{e} polynomials of quiver representation spaces. A K\"{a}hler manifold is endowed with a natural $Sl_2$ Lefschetz action on the cohomology. And the generating function of the Euler numbers mentioned above captures this action because Euler numbers are, after all, characterized by the cohomology. However, the refined generating function of the Betti numbers exactly organizes BPS states into different representations of the Lefschetz action. States invariant under this action have been conjectured to be special, in that they are expected to capture single-center black hole entropy. With the results presented in this article, one may identify the Lefschetz singlets in the low-lying non-polar terms to test the conjecture of \cite{Bena:2012hf,Lee:2012sc} that Pure-Higgs states make up single-center indices in all the above examples. The Poincar\'{e} polynomials studied in the mathematics literature that encode these invariants are based on Reinike's solution to the Harder-Narasimhan recursion for quivers without oriented closed loops. However, the indices used in this article, originally proposed in \cite{Manschot:2010xp,Manschot:2013dua}, applies to those with or without closed loops. It would be interesting to understand a mathematical counterpart of the latter, as extensions of Reinike's results. On the other hand, one may seek to understand a pattern of growth of multi-center entropies to compare against asymptotic behaviour of states under various representations that has been predicted in \cite{Bringmann:2013yta}.

\section*{Acknowledgements}
I thank Jan Manschot for introducing me to various problems in this field of research. Furthermore, I thank him for extended help during various stages of my learning of several aspects relevant to this work; in addition, I also thank him for some incisive comments during the initial stages of this work. I thank Jan Manschot and Stefan Vandoren for comments on a draft of this article. I thank Janu Verma for a guide into some relevant math literature. I was supported by a scholarship of the Bonn-Cologne Graduate School (BCGS) when this work was originally initiated. This work is now supported by the Netherlands Organisation for Scientific Research (NWO) under the VICI grant 680-47-603, and the Delta-Institute for Theoretical Physics (D-ITP) that is funded by the Dutch Ministry of Education, Culture and Science (OCW).

\clearpage

\appendix

\section{Three-center $D6$-$\bar{D6}$-$D2_{D0}$ configurations in the quintic}
The allowed $D2$-$D0$ charges for a generic $D6$-$\bar{D6}$-$D2_{D0}$ three-center configuration to have a non-vanishing index are collected in \autoref{tab:D2splitflowtable} below. 
\begin{table}[h!]
\begin{center}
\begin{tabular}{|c|c|}
  \hline
  $q_0$ & $q$ \\
  \hline \hline
  -4 & None\\
  \hline
  -3 & None\\
  \hline
  -2 & None\\
  \hline
  -1 & None\\
  \hline 
  0 & None\\
  \hline
  1 & 0\\
  \hline
  2 & -1, 0\\
  \hline
  3 & -2, -1, 0\\
  \hline 
  4 & -3, -2, -1, 0\\
  \hline
  5 & -4, -3, -2, -1, 0\\
  \hline
  6 & -5, -4, -3, -2, -1, 0\\
  \hline
  7 & -6, -5, -4, -3, -2, -1, 0\\
  \hline 
  8 & -7, -6, -5, -4, -3, -2, -1, 0\\
  \hline 
  9 & -8, -7, -6, -5, -4, -3, -2, -1, 0\\
  \hline
  10 & -9, -8, -7, -6, -5, -4, -3, -2, -1, 0, 1\\
  \hline
  11 & -10, -9, -8, -7, -6, -5, -4, -3, -2, -1, 0, 1\\
  \hline
  12 & -11, -10, -9, -8, -7, -6, -5, -4, -3, -2, -1, 0, 1\\
  \hline 
  13 & -12, -11, -10, -9, -8, -7, -6, -5, -4, -3, -2, -1, 0, 1\\
  \hline
  14 & -13, -12, -11, -10, -9, -8, -7, -6, -5, -4, -3, -2, -1, 0, 1\\
  \hline
  15 & -14, -13, -12, -11, -10, -9, -8, -7, -6, -5, -4, -3, -2, -1, 0, 1\\
  \hline
  16 & -15, -14, -13, -12, -11, -10, -9, -8, -7, -6, -5, -4, -3, -2, -1, 0, 1\\
  \hline
  17 & -16, -15, -14, -13, -12, -11, -10, -9, -8, -7, -6, -5, -4, -3, -2, -1, 0, 1, 2\\
  \hline
  18 & -17, -16, -15, -14, -13, -12, -11, -10, -9, -8, -7, -6, -5, -4, -3, -2, -1, 0, 1, 2\\
  \hline
  19 & -18, -17, -16, -15, -14, -13, -12, -11, -10, -9, -8, -7, -6, -5, -4, -3, -2, -1, 0, 1, 2\\
  \hline
\end{tabular}
  \caption{Allowed values for $q$ and $q_0$.}
  \label{tab:D2splitflowtable}
\end{center}
\end{table}
None of these allowed values satisfy regularity, proving the non-existence of the corresponding three-center solutions.

\clearpage

%
%
\bibliography{EllipticGenera_Multicenters}{}
\bibliographystyle{ieeetr}

\end{document}